\documentclass[aps,preprint,showpacs,showkeys,amsmath,amssymb,superscriptaddress,groupedaddress,showkeys]{revtex4}  
\usepackage{graphicx}  
\usepackage{subfigure}
\usepackage{dcolumn}   
\usepackage{bm}        
\usepackage{amssymb}   
\usepackage{multirow}
\usepackage{indentfirst}
\usepackage{subfigure}
\usepackage{array}
\usepackage{xcolor}
\usepackage{comment}

\newcommand{\PBUITEMS}{\affiliation{Department of Physics, Balochistan University of Information Technology, Engineering \& Management Sciences (BUITEMS),  Quetta, Pakistan}}
\newcommand{\CBUITMES}   {\affiliation{Department of chemistry, BUITEMS, Quetta , Pakistan}}
\newcommand{\EBUITMES}{\affiliation{Department of Environmental Science, BUITEMS, Quetta, Pakistan}}
\newcommand{\Texas}{\affiliation{The University of Texas MD Anderson Cancer Center, Houston, TX, USA.}}
\newcommand{\GUni}{\affiliation{Department of Physics, Hafiz Hayat Campus University of Gujrat, Gujrat, Pakistan}}

\begin{document}

\title{Dose Verification of Volumetric Modulated Arc Therapy using One and Two Dimensional Dosimeters
}

\author{Jalil ur Rehman}\PBUITEMS\GUni
\author{Zahra}\PBUITEMS
\author{Ghulam Hussain\footnote{Corresponding author: gh\_tuhep@hotmail.com}}\PBUITEMS
\author{Nisar Ahmad}\PBUITEMS
\author{H M Noor ul Huda Khan Asghar}\PBUITEMS
\author{Zaheer Abbas Gilani}\PBUITEMS
\author{Gulfam Nasar}\CBUITMES
\author{Malik M Akhter}\EBUITMES
\author{Muhammad Isa}\GUni
\author{Geoffrey S. Ibbott}\Texas

\date{\today}

\begin{abstract}
\textbf{Purpose}: To verify dose delivery and quality assurance of volumetric modulated arc therapy (VMAT) for head and neck cancer.

\textbf{Method}: The Imaging and Radiation Oncology Core Houston (IROC-H) head and neck phantom with thermo- luminescent dosimeters (TLDs) and films, were imaged with computed tomography scan and the reconstructed image was transferred to pinnacle treatment planning system (TPS). On TPS the planning target volume (PTV), secondary target volume (STV) and organ at risk (OAR) were delineated manually and a treatment plan was made. The dose constraints were determined for the concerned organs according to IROC-H prescription. The treatment plan was optimized using adoptive convolution algorithm to improve dose homogeneity and conformity. The dose calculation was performed using C.C Convolution algorithm and a Varian True Beam linear accelerator was used to deliver the treatment plan to the head and neck phantom. The delivered radiation dose to the phantom was measured through TLDs and GafChromic EBT2 films. The dosimetric performance of the VMAT delivery was studied by analyzing percent dose difference, iso-dose line profile and gamma analysis of the TPS computed dose and linac delivered doses.

\textbf{Result}: the percent dose difference of 3.8\% was observed between the planned and measured doses of TLDs and a 1.5mm distance to agreement (DTA) was observed by comparing iso-dose line profiles. Passed the gamma criteria of 3\%/3 mm was with good percentages. 

\textbf{Conclusion}: The dosimetric performance of VMAT delivery for a challenging head and neck radiotherapy can be verified using TLDs and films imbedded in an anthropomorphic H\&N phantom.

\end{abstract}


\maketitle

\section{Introduction}
The use of radiotherapy is mostly appreciated for head and neck (H\&N) cancer\cite{1}, but it also possesses many challenges. Mostly the H\&N tumor is in close proximity to critical organs therefore saving these organs from intolerable radiation exposure is a bit difficult. H\&N radiotherapy has been improved with the introduction of intensity modulated radiotherapy (IMRT)[2]. IMRT is usually used for H\&N cancer but its main drawbacks are the delivery of treatment in a longer time and involvement of larger number of monitor units (MUs)[3, 4], to overcome these limitations VMAT has been introduced.
VMAT is the new emerging technology whose supremacy is evidenced for various cancer sites including prostate, head and neck, lungs and breast etc[5-9]. VMAT is an arc based Intensity Modulated Radiotherapy(IMRT) having the capability of modulating gantry rotation speed in addition to variable dose rate and changing treatment aperture shapes[10]. It is the advanced form of IMRT that can deliver the whole treatment within few arcs with the reduced treatment time [11]. Short treatment time has the benefit of improving patient comfort, reduced intra-fraction motion and enhances overall treatment outcomes. 
VMAT has the superiority compared to other arc based techniques that it can be delivered with a conventional linear accelerator. VMAT optimization algorithm is the cause of its eminence[11], that optimize the dose rate, gantry speed and MLC positions in a single arc until the desired dose distribution is achieved. The optimization algorithm starts with a limited number of inputs gradually increases to a large number to give the planned dose profile[12, 13]. Another potential advantage of VMAT is the involvement of lower number of monitor units (MUs) which significantly lessens the integral dose to the patient thus decreasing the risk of secondary cancer[14].
VMAT is a complex treatment modality and its clinical implementation for head and neck cancer requires very accurate acceptance testing, commissioning of delivery and treatment planning system (TPS) and a very comprehensive quality assurance program. Patient specific quality assurance is a core element of quality assurance procedure, which is emphasized by most of the professional institutionsAmerican Association of Physicists in Medicine(AAPM), American College of Radiology (ACR), American Society for Radiation Oncology (ASTRO)[15, 16]. Patient specific QA is a pre-treatment validation, which is performed to check the accuracy (feasibility) of the planned treatment. Pretreatment validation is often marked with better radiotherapy results that ensures the delivery of the planned dose to the patient[17, 18]. Various reports and articles of AAPM are published to provide a direction to perform the patient specific QA.
This article has been presented to experimentally verify the dosimetric performance of clinical VMAT through one dimensional and two dimensional dosimeters in an Imaging and Radiation Oncology Core Houston (IROC-H) anthropomorphic H\&N phantom. The IROC-H phantom is a well known and standard phantom, used for credentialing of institutions.

\section{MATERIALS AND METHODS}
\subsection{IROC head and neck phantom}
The IROC anthropomorphic head and neck (H\&N) phantom, made up of an outer plastic shell with water tight polystyrene insert in it. The plastic shell is filled with water to substitute human tissue and the insert was designed to resemble the oropharynx, related lymph nodes and spinal cord. The oropharynx was decided to be our primary planning target volume (PTV) and secondary planning target volume (STV) was the lymph nodes and spinal cord was determined to be organ at risk (OAR). The plastic insert with its target structures and OAR are imageable being constructed of a material having different density than surrounding water and plastic shell. The insert had slits and holes to hold radio-chromic films and thermoluminescent dosimeters (TLDs).

\subsection{Accommodation of TLDs and films in phantom}
Capsulated TLDs were used to measure delivered point doses and radio-chromic EBT2 films [GafChromicVR EBT2, International Specialty Products (ISP), NJ] were used for the measurement of planar dose distribution. Total eight TLDs were placed in insets, four in primary PTV region, two in STV and two in OAR regions. Sheets of EBT2 film were put at the central axial planes of STV and OAR and one EBT2 film piece was placed at the primary PTV in the sagittal plane. The IROC H\&N phantom with the insert, TLDs and film is shown in Figure 1.
\begin{figure}
\includegraphics[width=12.27cm, height=10.82cm]{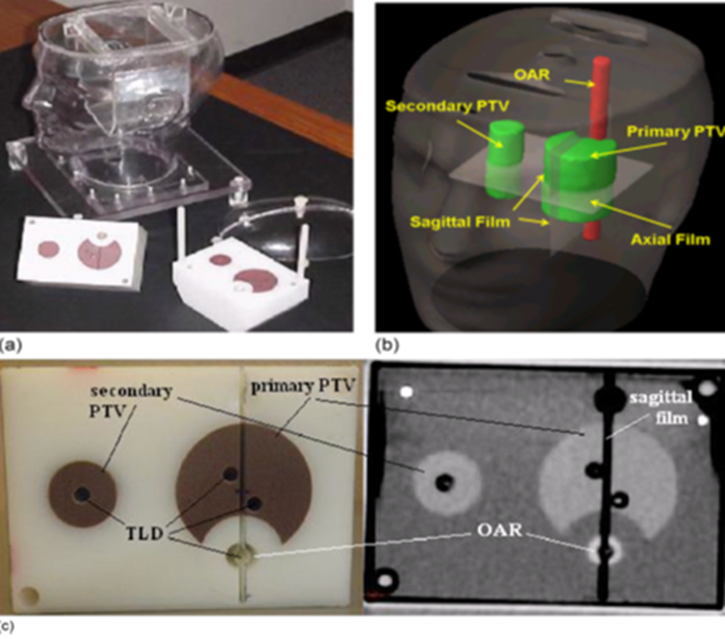}
\caption{IROC-H head and neck IMRT phantom (a) and inserts: PTV, STV, OAR, three films (b) and eight TLD capsules. The location of four TLD capsules in the upper part of the insert (c) and another set of four TLD capsules were in the corresponding location in the lower part of the insert; it also shows one axial CT slice of the inserts.}
\label{fig:rawsamp}
\end{figure}

\subsection{Treatment planning and delivery}
The computed tomography (CT) slice of 1.5mm thickness of IROC H\&N phantom was obtained by Philips CT scanner (Philips Healthcare, Andover, MA). The CT images were imported through Digital Imaging and Communications in Medicine (DICOM) to a pinnacle treatment planning system (TPS) for segmentation and planning. The contouring of all the structures and TLDs were done manually on TPS and then the procedure of treatment planning started. A dose criteria 0f 6.6 Gy and 5.4 Gy were prescribed for 95\% volume of primary PTV and STV, and maximum dose limit of 4.5 Gy was set for OAR. Four arcs of 182o-178o, 180o-184o, 182o-178o, 180o-184o were planned for VMAT delivery and the plan was optimized by adaptive convolve algorithm until the set goals were achieved. The VMAT plan geometry is shown in Figure 2, where dose distribution for different planes has also been shown. The doses to be delivered were calculated using C.C convolution algorithm.
The phantom was laid on the couch of a Varian True beam (Varian Medical Systems, Palo Alto, CA), linear accelerator, to deliver the planned treatment, the delivery of photons was repeated three times for the plan reproducibility.
\begin{figure}
\includegraphics[width=16.48cm, height=4.29cm]{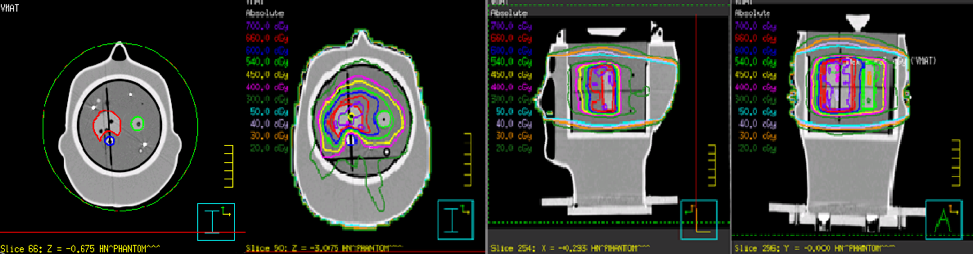}
\caption{VMAT plan geometry, dose distribution in the axial, sagittal and coronal view of the phantom containing film from left to right.}
\label{fig:rawsamp}
\end{figure}

\section{DOSE MEASUREMENTS AND ANALYSIS
}
Three types of test were performed to see the agreement between the planned and delivered doses to the phantom, given as: TLD Percent dose difference, Iso-dose line profile and Gamma analysis. TLD measured doses were compared with the mean computed dose of TLD calculated from TPS contours, and percentage difference were evaluated between the two. For film analysis the doses from three tiny pinholes drilled in each film were compared with the planned dose. Pixel resolution of 0.3mm was set for film study. The film dose was normalized by taking average of corresponding superior and inferior TLD doses. The planar dose distribution was measured from two cross profiles constructed on each film representing a sagittal or axial plans.

Gamma index analysis was performed to find the agreement between the computed dose from TPS and recorded doses from film. The 3\% dose difference and 3-mm distance-to-agreement (DTA) criteria were set to analyze the difference between planned and measured doses. Whereas criteria of 7\%/4 mm were used in IROC for the purpose of quality assurance of IMRT and VMAT. Iso-dose line profiles were also constructed to visualize the agreement between computed and planned dose. All the analysis and graph preparation was performed by Matlab R2008b (The MathWorks,Inc., Natick, MA) software.

\section{RESULTS}
\textbf{TLD result}

The comparison between the computed and measured point doses for all the eight TLDs are shown in table 1. In this table the standard deviation (σ) of the three deliveries is calculated which was less than 4\% and relative standard error (RSE) was also evaluated that gave the values within 2\%. The percentage dose difference evaluation of planned and delivered TLD dose was found to be within 5\% in all the eight locations where TLDs were inserted. The average percentage difference of 3.8 \% was obtained and this gradient reduced to 3.4\% when OAR percentage difference was excluded.   

\begin{table}
 \footnotesize
\begin{tabular*}{170mm}{@{\extracolsep{\fill}}ccccccccc}
\toprule 
TLD position & \multicolumn{3}{c}{Measurement}  & Average   & Std. deviation  & RSE$(\%)^a$& Planned~TLD  & Percentage~Diff. \\
		&\rm$T_1$& $T_2$ & $T_3$ &&&&& \\ 
				& cGy& cGy & cGy &cGy&$\sigma$&&cGy&$(\Delta\%)^b$ \\ \hline\hline

PTV S-A	& 663.5  & 666.5   &        662.5 &          664.2   &   1.6997      &0.2559 &   702.8 & 5.8168\\
PTV S-P	& 670.5  &  671.2  &        668.3  &         670  	&           1.2356 & 0.1844&  701.7  & 4.7313\\
PTV I-A	& 651.9  &  660.4  &        659.3  &         657.2	&           3.7745 & 0.5743&  674.9  & 2.6932\\
PTV I-P	& 653.7  &  662.6  &        655   	 &         657.1	&           3.9251 & 0.5973&  676.8  & 2.9980\\
STV-S  	& 558.9  &  559.1  &        561.4  &         559.8	&           1.1343 & 0.2026&  568    & 1.4648\\
STV-I  	& 556.5  &  557.6  &        549.4  &         554.5	&           3.6341 & 0.6554&  569.1  & 2.6330\\
CORD-S 	& 295.8  &  301.1  &        301.6  &         299.5	&           2.6242 & 0.8762&  315.7  & 5.4090\\
CORD-I 	& 294.7  &  298.1  &        302.4  &         298.4	&           3.1507 & 1.0558&  312.8  & 4.8257\\
Average &&&&&&&                                                                            			& 3.8214\\
$\Delta$\%&&&&&&&                                                                                           								&Except\\
&&&&&&&                                                                                          								 & cord\\
&&&&&&&                                                                                          								 &3.38953\\								 
 \hline
\end{tabular*}
\caption{\label{tab:GDMStable}
The comparison between TLD measurements and the dose calculations from treatment planning system (Pinnacle@3).}
\end{table}

\textbf{Film result}
The comparison between pinnacle TPS and EBT2 film in the form of line profiles only for one irradiation, and similar results were observed for the other two deliveries. The line profile comparison of the delivered and planned doses exhibit a good agreement and the slight discrepancies which were seen they were also within the tolerance limit of 4mm, provided by IROC as shown in figure 3. Maximum disagreements were present at the edges of the line profiles and film doses were bit more rough than the planned TPS dose profile.

\begin{figure}
\includegraphics[width=16.5cm,height=4.76cm]{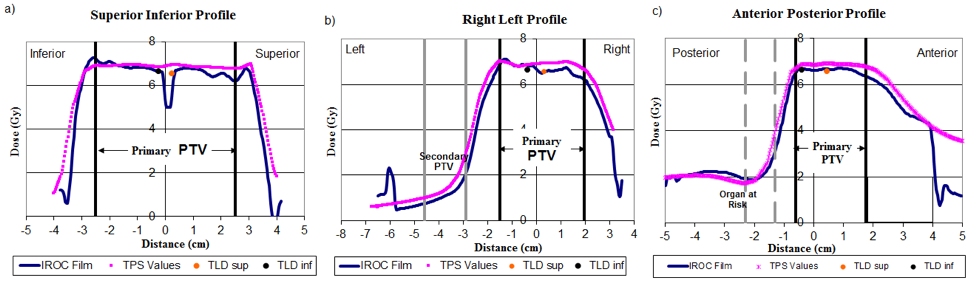}
\caption{Iso-dose line profile comparison of EBT2 film and pinnacle TPS in inferior-superior, left-right, posterior-anterior for first delivery. 
}
\label{fig:GDMS-profile}
\end{figure}

\textbf{Gamma analysis}
To further authenticate the delivered doses gamma analysis was performed which gives the dose difference and distance to agreement test in a single analysis. Figure 4 shows the gamma analysis performed for one treatment delivery which is representative of all the three irradiations. It was observed that the test passed the criteria of 3\%/3mm used by IROC for the purpose of credentialing of the institutions. The passing percentage for gamma analysis of the three irradiation is given in Table 2 below

\begin{table}
 \footnotesize
\begin{tabular*}{73mm}{@{\extracolsep{\fill}}ccccc}
\toprule 
Film position & \multicolumn{4}{c}{VMAT (Pixel\%)}\\
&$T_1$&$T_2$&$T_3$&Average\\\hline\hline
Axial  &98.50 &99.40&98.25&98.72\\
Sagittal&99.50&98.60&96.30&98.13\\ 
Axial PTV&99.70&97.80&98.20&98.56\\
   \hline
\end{tabular*}
\caption{\label{tab:GDMStable}
Percentage of points passing gamma analysis of 3\%/3mm criteria for all the three deliveries).}
\end{table}

\begin{figure}
\includegraphics[width=15.81cm, height= 5.01cm]{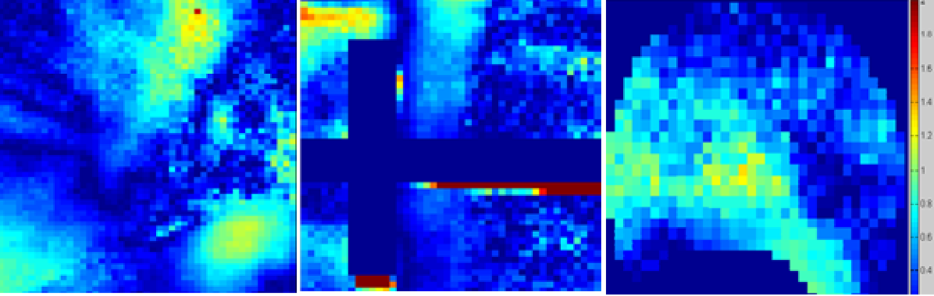}
\caption{Gamma map analysis of EBT2 film and TPS in the axial, sagittal and PTV planes for 3\%/3mm criteria
}
\label{fig:GDMS-profile}
\end{figure}

\section{DISCUSSION}
In this research study VMAT treatment planning for H\&N cancer has been verified using IROC anthropomorphic phantom with one-dimensional and two-dimensional dosimeters. The agreement between the planned and measured doses has been studied through evaluating percent dose difference, constructing iso-dose line profile and analyzing the 2D gamma maps. The IROC passing criteria of 3\% dose difference and a DTA of 3mm is used within PTV regions. The OAR region has been excluded from analysis being very sensitive to error. 
TLD analysis have given a dose difference of 3.3\% without spinal cord and 3.8\% with spinal cord, and a DTA of 1.5 mm has been observed between planned and measured iso-dose line profiles. The gamma analysis has passed the criteria of 3\%/3mm with average passing pixels of more than 98\%. The disagreement in all cases are within the acceptable limit, therefore VMAT can be delivered to patient with better target coverage and better organ sparing.
Delivering planned VMAT dose to head and neck cancer site is a challenging task, except the phantom setup error, limitation of dosimetric tools, contouring challenges, VMAT planning and delivery itself is a complex procedure. The Planning VMAT delivery taking into account the MLCs penumbra and leakage radiation, its tongue and groove effect, linac head scattering are very complicated. Calculating dose for such an uncertain system in addition to considering patient heterogeneities makes VMAT process inaccurate[19]. 
In VMAT many parameters have to be modulated, and significant dosimetric errors have been evidenced due to continuous variation of gantry speed and MLC shape [20]. Another problem is a heavy load of treatment machine head which can't be changed always , therefore sometimes larger standard deviations are observed in treatment deliveries, as it can be seen in this experiment in Table 1 as well. 
 MLC acceleration during irradiation also reduces VMAT accuracy and the greater speed of MLC is mostly accompanied with increasing errors in delivering the planned doses [21].
Film and TLD's energy dependence, fading factor, inaccurate film processing and TLD readout methods also reduces the accuracy of measurement of radiation doses[22, 23].
Our VMAT QA has successfully attained the IROC established criteria despite these all sources of errors, therefore VMAT can be delivered safely to a head and neck cancer patient.

\section{CONCLUSION}
VMAT is an advanced modality of delivering external beam radiotherapy with the potential advantage of short treatment time. Head and neck cancer radiotherapy which is quite challenging can be benefited by VMAT. The dosimetric performance of these complex procedures can be verified using anthropomorphic head and neck phantom with TLDs and films.

\section{Acknowledgements}

\end{document}